\documentclass[]{aastex701}
\usepackage{stix2}
\usepackage{amsmath,amsfonts,amssymb,cancel}
\usepackage{fix-cm}
\graphicspath{{./figures/}}

\newcommand{\lapprox} {\, \lower3pt\hbox{$\sim$}\llap{\raise2pt\hbox{$<$}}\,}
\newcommand{\gapprox} {\, \lower3pt\hbox{$\sim$}\llap{\raise2pt\hbox{$>$}}\,}

\shorttitle{Electron Transport and Heating in Solar Flares}
\shortauthors{Emslie \& Kontar}

\begin{document}

\title{Energy Transport and Heating by Non-Thermal Electrons in a Turbulent Solar Flare Environment}

\author[0000-0001-8720-0723]{A. Gordon Emslie}
\affiliation{Department of Physics \& Astronomy, Western Kentucky University, Bowling Green, KY 42101, USA}
\email[show]{gordon.emslie@wku.edu} 

\author[0000-0002-8078-0902]{Eduard P. Kontar}
\affiliation{School of Physics \& Astronomy, University of Glasgow, Glasgow, G12 8QQ, UK}
\email[show]{eduard@astro.gla.ac.uk} 

\begin{abstract}

The impulsive phase of a solar flare is known to generate strong turbulence and to transfer magnetic energy into accelerated electrons. Recognizing the importance of angular diffusion on the dynamics of the accelerated electrons, we extend previous treatments by deriving analytic solutions for the electron flux and associated energy deposition in two regimes: scattering dominated by inelastic Coulomb collisions and scattering dominated by elastic interactions with turbulent scattering centers. We show that the turbulence-dominated scattering term strongly reshapes the spatial distribution of the plasma heating: compared to the traditional collisional thick-target approach, turbulent scattering could lead to an order-of-magnitude increase in coronal heating and an even greater suppression of chromospheric heating. Scattering also acts to reduce the anisotropy of the electron distribution and so reduces the net current associated with the nonthermal electrons. The return-current Ohmic heating is accordingly reduced to a level that renders it negligible compared to direct collisional heating. The results have significant implications for models of atmospheric response to impulsive phase energy release, in particular chromospheric evaporation, flare-driven coronal heating, the formation of loop-top hard X-ray sources, and the longstanding discrepancy between modeled and observed soft X-ray line profiles.

\end{abstract}


\section{Introduction}\label{sec:intro}

There is growing evidence that a high level of plasma turbulence is created during the impulsive-phase energy release of a solar flare. Turbulent flows are a natural consequence of the high Reynolds number in the solar corona \citep[e.g.,][]{2014masu.book.....P}, and turbulence is often invoked as a means to enhance the plasma resistivity to the extent necessary to drive magnetic reconnection on timescales compatible with those of the impulsive phase of a flare \citep[see, e.g.,][]{1971ApJ...169..379C}. As the plasma turbulence is spatially and temporally well‑correlated with the power in accelerated electrons, it suggests that turbulence itself plays an active role in the flare energy‑transfer process \citep{2017PhRvL.118o5101K,2021ApJ...923...40S}.

\citet{2018ApJ...852..127B} have shown that suppression of heat conduction by turbulence can help explain the relatively long cooling times that are ubiquitously observed in post-flare loops \citep{1980sfsl.work..341M,2013ApJ...778...68R,2015A&A...584A..89J}, thus reducing the level of heating previously thought to be required in the post-impulsive phase of a flare. \cite{2022ApJ...939...19E} \citep[see also][]{2024ApJ...965....2D} showed that a turbulence-dominated mean free path in the corona leads to a weaker temperature dependence of the thermal conduction coefficient \citep[from its $T^{5/2}$ behavior for collisional transport;][]{1962pfig.book.....S} and so to shallower temperature gradients in the relatively cool lower corona and hence a greater amount of material at temperatures $\lapprox 10^6$~K. The correspondingly enhanced differential emission measure at such levels in the atmosphere offers a possible resolution to a long-standing enigma \citep[see, e.g.,][and references therein]{2023ApJ...943..120G} regarding the discrepancy between modeling results and the observationally-inferred behavior of the differential emission measure at low coronal temperatures.

EUV and soft X ray spectral lines observed during flares frequently exhibit widths that are significantly in excess of the thermal Doppler width \citep[e.g.,][]{1982SoPh...78..107A,1990A&A...236L...9A,1992str..book.....M,2001ApJ...549L.245H,2010A&A...521A..51P,2015ApJ...799L..12D}, strongly indicating the presence of some form of turbulence.  Indeed, \cite{2017PhRvL.118o5101K} have shown that high levels of turbulence, as evidenced by excess broadening of soft X-ray spectral lines, are present in coronal regions where flare energy release occurs, and they further showed that the energy content in the turbulence, coupled with the relatively short timescales over which turbulence is created and dissipated, makes turbulent motions a very effective conduit of energy from the chaotic dissipation of magnetic field in the region of primary energy release to other forms of energy \citep[e.g., non-thermal electrons, heating, mass motions;][]{2012ApJ...759...71E} that drive the flare. \citet{2021ApJ...923...40S} showed that turbulence can be distributed throughout the flaring region, and is often greatest in the coronal loop tops, where higher energy electrons suffer stronger scattering \citep{2018A&A...610A...6M}.

Several studies \citep{2014ApJ...780..176K,2016ApJ...824...78B,2018ApJ...852..127B,2018ApJ...865...67E,2022ApJ...931...60A} have considered the effects of turbulent scattering on energy transport by thermal conduction, and recently \citep{2024ApJ...977..246E,2025ApJ...993..127E} the effects of turbulent scattering on the electrical resistivity of the ambient medium, and so on the Ohmic heating effected by beam-neutralizing return currents \citep{1977ApJ...218..306K,1980ApJ...235.1055E,1984A&A...131L..11B,1989SoPh..120..343L,1990A&A...234..496V,1995A&A...304..284Z,2006ApJ...651..553Z,2017ApJ...851...78A} has also been considered.

It is widely accepted \citep[e.g.,][]{2012ApJ...759...71E,2016ApJ...832...27A,2019ApJ...881....1A} that a very significant component of the energy released during a solar flare appears in the form of a beam of nonthermal electrons, accelerated in the primary energy release site. Complete consistency of any treatment of impulsive-phase energy transport in a turbulent environment must, therefore, also consider the effects of turbulent scattering on the energy deposition rate effected by the passage of such a beam of nonthermal electrons through the ambient solar atmosphere. The transfer of energy from the non-thermal electrons to the ambient atmosphere proceeds through two main processes: direct collisions of beam electrons on ambient electrons and Ohmic heating associated with driving the beam-neutralizing return current through the finite resistivity of the ambient plasma. In a turbulent medium, the electron trajectories are also affected by scattering off turbulent inhomogeneities. Although such (elastic) collisions do not constitute an additional energy transfer mechanism \emph{per se} between the nonthermal electrons and the ambient solar atmosphere, they do alter the trajectories of the electrons, resulting in a modification of the spatio-spectral distribution of the electron spectrum and hence the associated heating rate \citep{2014ApJ...780..176K}.

The present work explores an electron transport model in which the energy loss still takes place through Coulomb collisions on ambient electrons and/or Ohmic heating associated with the driving of the beam-neutralizing return current, but in which the electron trajectories include the effects of scattering, either through inelastic Coulomb collisions or through elastic scattering off turbulent scattering centers. If the turbulence scattering length $\lambda_T$ is large compared to the collisional mean free path for an electron of a specified energy $E$, then turbulent scattering can be neglected; the scattering is then dominated by Coulomb collisions. In such a situation the collisional mean free path is somewhat less than the total travel distance and so the electrons propagate diffusively. This results \citep{2018ApJ...862..158E} in an energy deposition vs. depth profile that is rather similar to that in traditional non-diffusive treatments in which the electron transport is modeled by computing the dynamics of scatter-free particles
\citep[][]{1984ApJ...279..896N,1989ApJ...341.1067M,2005ApJ...630..573A,2009ApJ...702.1553L,2022ApJ...931...60A}.
For field-aligned injection \citep[Figure~1 of][]{2018ApJ...862..158E} the heating profiles in a collisional diffusive model are similar in magnitude to those calculated using a non-diffusive model, but in general vary more smoothly with distance from the injection site; the local heating rate differs from that of the non-diffusive model by at most a factor $\sim$$2$. For isotropic injection, the differences between the diffusive and non-diffusive treatments are even smaller; the diffusive heating profiles are very similar to the non-diffusive results \citep[Figure~1 of][]{2018ApJ...862..158E}.

More significant differences in the heating profile are to be expected in scenarios in which $\lambda_T < \lambda_C$, so that the scattering is now controlled by the turbulence scattering scale. In such a scenario, the value of the scattering mean free path $\lambda_T$ can be observationally constrained: for example, in their study of electron acceleration in solar flares, \cite{2021ApJ...923...40S} show that a turbulent scattering length $\lambda_T \simeq 2 \times 10^8$~cm provides the best fit to the accelerated electron spectrum, as determined from various aspects (e.g., spectrum, spatial distribution) of the emitted hard X-rays. For comparison, in the flaring corona, with $T \simeq 3 \times 10^7$~K and $n \simeq 3 \times 10^{10}$~cm$^{-3}$, the collisional mean free path $\lambda_C \simeq 10^9$~cm, so that it appears that electron acceleration proceeds in an environment in which the ratio $\lambda_T/\lambda_C \lapprox 1$. Similarly, in their study of electron transport and energy loss in the surrounding target atmosphere, \cite{2014ApJ...780..176K} conclude that while there is evidence to support a scattering mean free path that is less than the collisional value, ``models that invoke mean free paths smaller than $\sim 10^8$~cm (or equivalently electron isotropization times shorter than $10^{-2}$~s) are difficult to reconcile with the data'': if $\lambda_T/\lambda_C \ll 1$, then turbulent scattering becomes so dominant that the variation of the electron distribution in space is manifestly inconsistent with observations.

\cite{2018A&A...610A...6M}, in their analysis of radio and X-ray emissions, deduce similar values for $\lambda_T$ in the upper corona. Further, in a rather different context, \cite{2022ApJ...939...19E} show that a $\lambda_T$ of $10^6 - 10^7$~cm is necessary to reduce temperature gradients in the low corona of active region loops to a level where the associated differential emission measure has a value that is consistent with observations \citep{2023ApJ...943..120G}. For comparison, the collisional mean free path in the low corona ($T \simeq 3 \times 10^6$~K, $n \simeq 3 \times 10^{10}$~cm$^{-3}$) is of order $10^7$~cm, so that, once again, the ratio $\lambda_T/\lambda_C$ is a few tenths.

We thus see that there is considerable justification to explore situations where $\lambda_T/\lambda_C$ is $\sim$$(0.1-0.3)$. Indeed, such a range is the only one worth exploring: much larger values of this ratio imply that turbulent scattering can be neglected, so that a collisional treatment \citep{2018ApJ...862..158E} is adequate, whereas much smaller values of the ratio are inconsistent with available observational constraints. To illustrate the essential physics, we consider the case where $\lambda_T$ is constant throughout the target; this assumption is rigorously valid in the case of, e.g., scattering from magnetic inhomogeneities \citep[see the Appendix in][]{2014ApJ...780..176K}, and the results can readily be extended, via a numerical treatment \citep[e.g.,][]{2020ApJ...902...16A}, to situations where $\lambda_T$ is not uniform. The explored range of $\lambda_T/\lambda _C$ represents a domain in which the scattering and energy loss terms in the transport equation are both important, but are each governed by different physics; as we shall see, this leads to some rather intriguing results.

The outline of the paper is as follows. In Section~\ref{sec:analysis} we review the diffusive transport equation and present its solution in two cases: scattering dominated by (inelastic) Coulomb collisions, and scattering dominated by (elastic) scattering off turbulent magnetic fluctuations. Results are presented in Section~\ref{sec:results}; they show that, compared to the test-particle approach, or even a diffusive analysis characterized by a collisional mean free path \citep{2018ApJ...862..158E} the presence of turbulence results in very significant changes to the heating profile, in particular an enhancement of heating in the corona and a reduction in chromospheric heating. In Section~\ref{sec:discussion} we review the essence of the results obtained and their impact on a large variety of studies, and we also stress that for full consistency these results should be incorporated into numerical codes that study the response of a turbulent solar atmosphere to the injection of a beam of nonthermal electrons during the impulsive phase of a flare.

\section{Analysis}\label{sec:analysis}

\subsection{Solution to the electron transport equation in the diffusive regime}\label{diffusive-solutions}

Following ideas originally posited by \cite{2014ApJ...780..176K}, \cite{2018ApJ...862..158E} studied the transport of electrons, including the effects of \emph{angular} diffusion, but not the \citep[higher-order;][]{2017ApJ...835..262B} process of diffusion in \emph{energy}.  Equation~(6) of that paper shows that the electron flux spectrum $F(E,z)$ (electrons~cm$^{-2}$~s$^{-1}$) can effectively be modeled by the one-dimensional diffusive transport equation

\begin{equation}\label{eq:diffusion-continuity}
- \, \frac{\lambda}{6} \, \frac{\partial^2 F(E,z)}{\partial z^2} + \frac{\partial }{\partial E} \, \left [ \, B(E) \, F(E,z) \, \right ] =  S(E,z) \,\,\, ,
\end{equation}
where $\lambda$ is the scattering length (mean free path), $S (E,z)$ \citep[cm$^{-3}$~s$^{-1}$~erg$^{-1}$; $\hat{S}$ in the notation of][]{2018ApJ...862..158E} is a source function corresponding to the injection of electrons accelerated in the primary energy release site, and $B(E)$ (erg~cm$^{-1}$) is the usual \citep{1972SoPh...26..441B,1978ApJ...224..241E} cold-target energy loss rate (erg~cm$^{-1}$ per unit distance $z$):

\begin{equation}\label{eq:cold-target-be}
B(E) \equiv \left | \frac{dE}{dz} \right | = \frac{Kn}{E} \,\,\, .
\end{equation}
Here $K = 2 \pi e^4 \ln \Lambda \simeq 6.7 \times 10^{-36}$~erg$^2$~cm$^2$ is the usual collision parameter, $e$ (esu) being the electronic charge (esu) and $\ln \Lambda \simeq 20$ the Coulomb logarithm) and $n$ (cm$^{-3}$) is the ambient number density. For simplicity, we will adopt a uniform density $n$, so that the characteristic collisional stopping distance for an electron of injected energy $E$, in a scenario without diffusion, is $E^2/2Kn$.

The lead (diffusional) term in Equation~\eqref{eq:diffusion-continuity} is valid for any form of scattering, characterized only by the pertinent mean free path $\lambda$. The main purpose of the work by \cite{2018ApJ...862..158E} was to evaluate the effect of scattering in the collisional heating rate; hence the scattering was modeled self-consistently using a mean free path  $\lambda = \lambda_C = 2E^2/Kn$ appropriate for Coulomb collisions. Here we generalize this treatment to include\footnote{In general, the scattering mean free path $\lambda$ derives from a combination of collisional and turbulent scattering, with $\lambda^{-1} = \lambda_C^{-1} + \lambda_T^{-1}$, where the collisional mean free path is $\lambda_{C}(v) = v/\nu_{C}(v)$, with $\nu_C(v)$ the collision frequency for an electron of velocity $v$. Substituting $\nu_{C}(v) = 4\pi n_e \, e^4 \, \ln \Lambda/m_e^2 \, v^3$ \citep[e.g.,][]{1962pfig.book.....S} gives $\lambda_C = E^2/\pi n_e e^4 \ln \Lambda \equiv 2 E^2/Kn$. We treat the turbulence mean free path $\lambda_T$ as a constant, and assume that it is smaller than the collisional mean free path at all energies and densities of interest, so that $\lambda \simeq \lambda_T$.} turbulent scattering with an associated mean free path $\lambda_T$; however, we emphasize that because turbulent scattering is an elastic process, the energy loss rate $B(E)$ (erg~cm$^{-1}$) nevertheless still reflects only (inelastic) collisional processes. We consider two cases: one in which turbulent scattering is negligible compared to scattering by Coulomb collisions, and a second in which turbulent scattering dominates, so that $\lambda = \lambda_T$, which we take to be a constant.

Following \cite{2018ApJ...862..158E}, we introduce the new dependent variable

\begin{equation}\label{eq:Phi-def}
\Phi(E,z) = F(E,z) \, B(E) \equiv \frac{Kn}{E} \, F(E,z)
\end{equation}
(units cm$^{-3}$~s$^{-1}$) and a new independent energy variable (with units cm$^2$)

\begin{equation}\label{eq:zeta-def}
\zeta = \frac{1}{6} \, \int_E^{E_o} \frac{\lambda}{B(E)} \, dE \,\,\, ,
\end{equation}
where $E_o$ is the injected energy. The respective forms of the energy-related variable $\zeta$ in these two cases are 

\begin{equation}\label{eq:zeta_collisional-vs_turbulent}
\zeta =
\begin{cases}
\begin{aligned}
 \stackrel{\lambda =\lambda _C}{=}& \frac{1}{6Kn} \int_E^{E_o} \!\! \lambda_C(E') \, E' \, dE' \!\!\!\!\!&=& \frac{1}{3 (Kn)^2} \int_E^{E_{0}} \!\! E'^3 \, dE' \!\!\!\!\! &=& \, \ell_C^{2} \left [ \left ( \frac{E_{0}}{k_{B}T_{e}} \right )^{4} -
\left (\frac{E}{k_{B}T_{e}} \right )^{4} \right ] \!\!  &;& \, {\rm \, collisional\,scattering}
\cr
\stackrel{\lambda =\lambda _T}{=}& \frac{1}{6Kn} \int_E^{E_o} \!\! \lambda_T \, E' \, dE' &=& \frac{\lambda_T}{6 Kn} \int_E^{E_{0}} \!\! E' \, dE' \, &=& \, \ell_T^{2} \left [ \left ( \frac{E_{0}}{k_{B}T_{e}} \right )^{2} -
\left (\frac{E}{k_{B}T_{e}} \right )^{2} \right ] \!\!  &;&\, {\rm \, turbulent\, scattering} \,\,\, ,
\end{aligned}
\end{cases}
\end{equation}
where

\begin{equation}\label{eq:ell-def}
\ell_C = \frac{1}{\sqrt{12}} \frac{(k_B T_e)^2}{Kn} \equiv \frac{\lambda_{ec}}{4 \sqrt{3}} \, ; \qquad \ell_T = \frac{\sqrt{\lambda_T}}{\sqrt{12}} \frac{k_B T_e}{\sqrt{Kn}} \equiv \frac{\sqrt{\lambda_{ec} \, \lambda_T}}{2 \sqrt{6}} \,\,\, ,
\end{equation}
and the collisional mean  free path for an electron at the thermal speed is $\lambda_{ec} = 2 (k_B T_e)^2/K  n$. Note the appearance of both $\lambda_{ce}$ and $\lambda_T$ in the turbulent scattering case, since both collisional energy loss and turbulent scattering are involved in the electron dynamics.

In terms of the variables $\Phi$ and $\zeta$ given by Equations~\eqref{eq:Phi-def} and~\eqref{eq:zeta-def}, Equation~(\ref{eq:diffusion-continuity}) takes the form of a standard diffusion equation

\begin{equation}\label{eq:basic}
\frac{\partial \Phi}{\partial \zeta} = \frac{\partial^2 \Phi(\zeta,z)}{\partial z^2} + {\overline S}(\zeta,z) \,\,\, ,
\end{equation}
with $\zeta$ playing the role of time, a consequence of the monotonically decreasing behavior of the electron energy $E$. The quantity ${\overline S}(\zeta,z) = (6 B(E)/\lambda) \, S$ (cm$^{-5}$~s$^{-1}$) is the pertinent source function with respect to these variables, and we note for later that ${\overline S}(\zeta,z) \, d\zeta = S(E,z) \, dE$. This parabolic diffusion equation can be solved using the Green's function \citep[see Equation~(26) in][]{2014ApJ...780..176K}

\begin{equation}\label{eq:greens_function_Phi}
G_{\Phi}(\zeta,z; \zeta',z') = \frac{1}{\sqrt{4 \, \pi \, (\zeta'-\zeta)}} \, \exp \left ( - \frac{({z'-z})^{2}}{4 \, (\zeta'-\zeta)} \right ) \, H(\zeta' - \zeta) \,\,\, ,
\end{equation}
where $H(\cdot)$ is the Heaviside step function, so that the solution is finite only for $\zeta' > \zeta$, a consequence of the collisional stopping depth due to energy loss associated with Coulomb collisions. The resulting solution for $\Phi(\zeta,z)$ is

\begin{eqnarray}\label{eq:Phi_solution}
\Phi(\zeta,z) &=& 
\int_{z' = -\infty}^{+\infty} \int_{\zeta'=0}^\infty  \,\, 
{\overline S}(\zeta',z') \, G_{\Phi}(\zeta,z; \zeta',z') \, d\zeta' \, dz' 
\cr
&=& \int_{z' = -\infty}^{+\infty} \int_{\zeta'=\zeta}^\infty  \,\, 
{\overline S}(\zeta',z') \, \frac{1}{\sqrt{4 \, \pi \, \zeta'}} \, \exp \left ( - \frac{({z'-z})^{2}}{4 \, \zeta'} \right )\, d\zeta' \, dz' \,\,\, .
\end{eqnarray}

We assume that electrons with an flux spectrum $F_o(E_o)$ [electrons~cm$^{-2}$~s$^{-1}$~erg$^{-1}$] are injected at the point $z=0$. The source term in $(E,z)$ space is thus $S(E,z) = \delta(z) \, F_o(E_o)$, where $\delta (\cdot)$ is the Dirac delta-function. 
Using ${\overline S}(\zeta',z') \, d\zeta' = S(E_o, z') \, dE_o = F_o(E_o) \, \delta(z')$ to replace the integral over~$\zeta'$ by an equivalent integral over~$E_o$, the solution to the diffusion equation~\eqref{eq:basic} in terms of the original variables $(E,z)$ and the flux $F(E,z) = (E/Kn) \, \Phi(\zeta[E],z)$ is

\begin{equation}\label{eq:f-result}
F(E,z) \! = \!
\begin{cases}
\begin{aligned}
\frac{E}{Kn} \! \int_{E}^{\infty} \!\!\!
\frac{F_o(E_o) \, dE_o} {\left \{ 4 \pi l_C^2 \left [ \left (\frac{E_o}{k_B T_e} \right )^4 \!\! - \!
\left ( \frac{E}{k_B T_e} \right )^4 \right ] \right \}^{1/2}} \, \exp \left \{ - \, \frac{z^2}{4 \, \ell_C^2 \left [ \left ( \frac{E_o}{k_BT_e} \right )^{4} \!\! -
\! \left ( \frac{E}{k_B T_e} \right )^4 \right ]} \right \} \, &; \, {\rm \, collisional\,scattering}
\cr
\frac{E}{Kn} \! \int_{E}^{\infty} \!\!\!
\frac{F_o(E_o) \, dE_o}{\left \{ 4 \pi \, l_T^2 \left [ \left (\frac{E_o}{k_B T_e} \right )^2 \!\! - \!
\left ( \frac{E}{k_{B}T_{e}} \right )^{2} \right ] \right \}^{1/2}} \, \exp \left \{ - \, \frac{z^2} {4 \, \ell_T^2  \left [ \left ( \frac{E_o}{k_B T_e} \right )^2 \!\! -
\! \left ( \frac{E}{k_B T_e} \right )^2 \right ] } \right \} \, &; \,  {\rm \, turbulent \, scattering} \, .
\end{aligned}
\end{cases}
\end{equation}

For definiteness, we next adopt a low-energy-truncated power-law form for the source (accelerated) electron spectrum:

\begin{equation}\label{eq:injected_spectrum}
F_{0}(E_0)=\frac{\dot{N}}{A} \, \frac{(\delta -1)}{E_{c}} \, \left ( \frac{E_o}{E_c} \right )^{-\delta} \, ; \qquad E_o \ge E_c \,\,\, ,
\end{equation}
where $A$ (cm$^2$) is the cross-sectional area of the flaring loop, so that the total injected rate (s$^{-1}$) is $\dot{N} = A \, \int_{E_{c}}^{\infty} F_0(E_0) \, dE_0$. With this, we obtain

\begin{equation}\label{eq:fez-result-collisional-scattering}
F_C(E,z) = \sqrt{\frac{3}{\pi}} \, \frac{\dot{N}}{A} \, \frac{(\delta-1)}{E_{c}} \times
\begin{cases}
\begin{aligned}
E \, \int_{E_c}^{\infty} dE_{0} \, \frac{(E_{0}/E_{c})^{-\delta}}{(E_0^4-E^4)^{1/2}} \, \exp \left \{ - \frac{3 (Knz)^2}{E_0^4-E^4} \right \} \, &; \,\, E < E_c
\cr
E \, \int_{E}^{\infty} dE_{0} \, \frac{(E_{0}/E_{c})^{-\delta}}{(E_0^4-E^4)^{1/2}} \, \exp \left \{ - \frac{3 (Knz)^2}{E_0^4-E^4} \right \} \, &; \,\, E \ge E_c
\end{aligned}
\end{cases}
\end{equation}
for collision-dominated scattering 
\citep[cf. Equation~(19) in ][]{2018ApJ...862..158E}, and

\begin{equation}\label{eq:fez-result-turbulent-scattering}
F_T(E,z) = \sqrt{\frac{3}{\pi}} \, \frac{\dot{N}}{A} \, \frac{(\delta-1)}{E_{c}} \times
\begin{cases}
\begin{aligned}
\frac{E}{\sqrt{K n \lambda_T}} \, \, \int_{E_c}^{\infty} dE_{0} \, \frac{(E_{0}/E_{c})^{-\delta}}{(E_0^2-E^2)^{1/2}} \, \exp \left \{ - \frac{3 Knz^2/\lambda_T}{E_0^2-E^2} \right \} \, &; \,\, E < E_c 
\cr
\frac{E}{\sqrt{K n \lambda_T}} \, \int_{E}^{\infty} dE_{0} \, \frac{(E_{0}/E_{c})^{-\delta}}{(E_0^2-E^2)^{1/2}} \, \exp \left \{ - \frac{3 Knz^2/\lambda_T}{E_0^2-E^2} \right \} \, &; \,\, E \ge E_c
\end{aligned}
\end{cases}
\end{equation}
for turbulence-dominated scattering.

\bigskip

\subsection{Direct collisional heating}\label{sec:collisional-heating}

With the forms of $F_C(E,z)$ and $F_T(E,z)$ (for collision-dominated, and turbulence-dominated, scattering, respectively) now determined, we can calculate the energy deposition profile due to Coulomb collisions in each scenario. (We remind the reader that the diffusion term relates to pitch angle scattering only; diffusion in energy is a higher order effect \citep{2017ApJ...835..262B}. Thus a cold-target energy loss rate $dE/dz = -Kn/E$ is still appropriate in each scenario.) Following \cite{2018ApJ...862..158E}, we use Equations~\eqref{eq:fez-result-collisional-scattering} and~\eqref{eq:fez-result-turbulent-scattering} and reverse the order of integration over $E$ and $E_0$ to determine the energy flux ${\cal F}(z) = \int_0^\infty E \, F(E,z) \, dE$
[erg~cm$^{-2}$~s$^{-1}$] at each position $z$:

\begin{eqnarray}\label{eq:f-diff}
{\cal F}(z) &=& \sqrt{\frac{3}{\pi}} \, \frac{\dot{N}}{A} \,
\frac{(\delta-1)}{E_{c}} \times \cr
&\times & \begin{cases}
\begin{aligned}
   \int_{E_0 = E_c}^\infty \left ( \frac{E_0}{E_c} \right )^{-\delta} \, dE_0 \, \int_{E = 0}^{E_0} \frac{E^2 \, dE}{(E_0^4-E^4)^{1/2}} \, \exp \left \{ - \frac{3 (Knz)^2}{E_0^4-E^4} \right \}  \, ; \, &{\rm \,\, collisional \, scattering}
  \cr
  \frac{1}{\sqrt{K \, n \, \lambda_T}} \, \int_{E_0 = E_c}^\infty \left ( \frac{E_0}{E_c} \right )^{-\delta} \, dE_0 \, \int_{E = 0}^{E_0} \frac{E^2 \, dE}{(E_0^2-E^2)^{1/2}} \, \exp \left \{ - \frac{3 Knz^2/\lambda_T}{E_0^2-E^2} \right \} \,  ; \, &{\rm \,\, turbulent \, scattering} \,\,\, ,
\end{aligned}
\end{cases}
\end{eqnarray}
and hence the heating rate $Q(z) = - d{\cal F}/dz$:

\begin{eqnarray}\label{eq:q-diff}
Q(z) &=& \sqrt{\frac{3}{\pi}} \, \frac{\dot{N}}{A} \,
\frac{(\delta-1)}{E_{c}} \times
\cr
&\times&\begin{cases}
\begin{aligned}
\left ( 6 K^2n^2 z \right ) \, \int_{E_0 = E_c}^\infty \left ( \frac{E_0}{E_c} \right )^{-\delta} dE_0 \int_{E = 0}^{E_0} \frac{E^2 \, dE}{(E_0^4-E^4)^{3/2}} \, \exp \left \{ - \frac{3 (Knz)^2}{E_0^4-E^4} \right \} \, ; \, &{\rm \,\, collisional \, scattering}
\cr
\left ( \frac{6 \sqrt{K n} \, z}{\lambda_T^{3/2}} \right ) \, \int_{E_0 = E_c}^\infty \left ( \frac{E_0}{E_c} \right )^{-\delta} dE_0 \int_{E = 0}^{E_0} \frac{E^2 \, dE}{(E_0^2-E^2)^{3/2}} \, \exp \left \{ - \frac{3 Knz^2/\lambda_T}{E_0^2-E^2} \right \}  \, ; \, & {\rm \,\,turbulent \, scattering}  \,\,\, . 
\end{aligned}
\end{cases}
\end{eqnarray}

\bigskip

\subsection{Return current Ohmic energy deposition}\label{return-current-ohmic}

Regardless of the degree of anisotropy in the injected electron distribution, electrons streaming from the acceleration region into opposite legs of the loop rapidly induce compensating return currents, carried by the thermal electrons in the target plasma, to maintain charge and current neutrality in each leg. Driving such return currents through the finite resistivity of the ambient plasma produces an associated Ohmic energy deposition rate

\begin{equation}
Q_{rc}(z) = j(z) \cdot {\cal E}(z) \,\,\, ,
\end{equation}
where the magnitude of the return current density $j$ is related to the beam spectral number density $n_b(E,z)$ (cm$^{-3}$) by

\begin{equation}
j (z) = e \, \int n_b (E,z) \, v(E) \, dE  = e  \int_0^\infty F(E,z) \, dE \,\,\, .
\end{equation}
For a local Ohm's law ${\cal E} = \eta \, j$, with scalar resistivity $\eta$, we thus have

\begin{equation}\label{q-rc-exp}
Q_{rc}(z) = \eta \, e^{2} \left ( \int_0^\infty F(E,z) \, dE  \right )^{2} \,\,\, .
\end{equation}
The form of $F(E,z)$ in this expression should, of course, be evaluated self-consistently using both collisional and return-current energy losses. However, as a first approximation, we use the (dominant, as will be justified {\it a posteriori} below) collisional diffusion result~\eqref{eq:fez-result-collisional-scattering} for $F(E,z)$, so that

\begin{eqnarray}\label{return-current-diffusion-result}
Q_{rc}(z) &=&\eta \, e^2 \, \left ( \frac{3}{\pi} \right ) \, (\delta-1)^2 \, \left ( \frac{{\dot N}}{A} \right )^2 \, \times \cr
&\times&
\begin{cases}
\begin{aligned}
  \left [ \frac{1}{E_c} \, \int_{E_0=E_c}^\infty \left ( \frac{E_0}{E_c} \right )^{-\delta} \, dE_0 \, \int_0^{E_0} \frac{E \, dE}{(E_0^4-E^4)^{1/2}} \exp \left \{ - \frac{3 (Knz)^2}{E_0^4-E^4} \right \} \right ]^2  \, &; {\rm \,\, collisional \, scattering}
\cr
  \frac{1}{K \, n \, \lambda_T} \, \left [ \frac{1}{E_c} \, \int_{E_0=E_c}^\infty \left ( \frac{E_0}{E_c} \right )^{-\delta} \, dE_0 \, \int_0^{E_0} \frac{E \, dE}{(E_0^2-E^2)^{1/2}} \exp \left \{ - \frac{3 Knz^2/\lambda_T}{E_0^2-E^2} \right \} \right ]^2 \, &;  {\rm \,\, turbulent \, scattering} \,\,\, .  
\end{aligned}
\end{cases}
\end{eqnarray}

To compute the heating rate, we used an ambient resistivity given by Equations~(5) and~(25) of \cite{2016ApJ...824...78B}, modified by the ratio of the collisional to turbulent mean free paths \citep[see][]{2024ApJ...977..246E}, viz.: 

\begin{equation}\label{eq:resistivity}
\eta = \frac{2 \pi e^2 \ln \Lambda \, m_e^{1/2}}{(2 k_B T_e)^{3/2}} \, \left ( \frac{\lambda_{ec}}{\lambda_T} \right ) \simeq 6 \times 10^{-9} \,\left ( \frac{\lambda_{ec}}{\lambda_T} \right ) \,  T_e^{-3/2} = 2 \times 10^{-19} \left ( \frac{\lambda_{ec}}{\lambda_T} \right ) \, {\rm s} \,\,\, ,
\end{equation}
where a temperature $T_e = 10^7$~K has been adopted in the last equality.

\section{Results}\label{sec:results}

In this section we present the results obtained using the analysis of the previous section. For definiteness, we show results for a particle flux ${\dot N} = 10^{18}$~cm$^{-2}$~s$^{-1}$ (corresponding to an energy flux of a few $\times \, 10^{10}$~erg~cm$^{-2}$~s$^{-1}$, or some $10^{29}$ erg~s$^{-1}$ for a flare area of $3 \times 10^{18}$~cm$^2$). The energy fluxes and collisional heating rates scale as ${\dot N}$, while the return current heating rate scales as ${\dot N}^2$. Results are shown for illustrative low-energy cutoffs $E_c = 10, 15,$ and $20$~keV, and spectral indices $\delta=4$ and $6$. To compute the collisional mean free path we use an ambient density $n = 3 \times 10^{10}$~cm$^{-3}$ and we further use an electron temperature $T_e = 10^7$~K to evaluate the resistivity used in computing the return current Ohmic heating. Together, these give a collisional mean free path (for an electron moving at the thermal velocity) of $7 \times 10^7$~cm, and we consider values of the turbulence scattering length equal to $0.1 \times, 0.2 \times$, and $0.3 \times$ this, i.e., $\lambda_T = 7 \times 10^6, 1.5 \times 10^7$, and $2 \times 10^7$~cm \citep[cf. the $10^6 - 10^7$~cm values used by][]{2022ApJ...939...19E}. We also use an illustrative loop half-length of $3 \times 10^9$~cm (40 arc seconds on the solar disk); this gives a coronal column density $N_{\rm cor}$ of $(3 \times 10^{10}) \times (3 \times 10^9) \simeq 10^{20}$~cm$^{-2}$, which is the collisional stopping distance for an electron of energy $E = \sqrt{2 K N_{\rm cor}} \simeq 25$~keV. The corona therefore acts as a thick target for electrons of energy less than this, while electrons of greater energy would, in a test-particle transport model, penetrate the coronal region of the loop and deposit the bulk of their energy at the loop footpoints. 

\begin{figure}[pht]
\begin{center}
\includegraphics[width=0.32\textwidth]{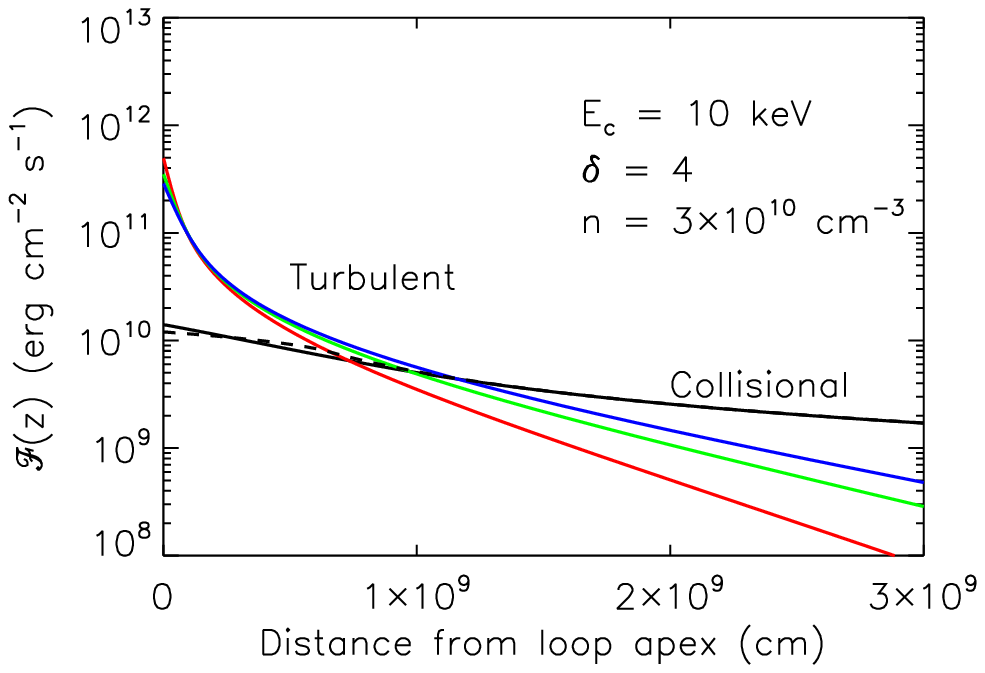}
\includegraphics[width=0.32\textwidth]{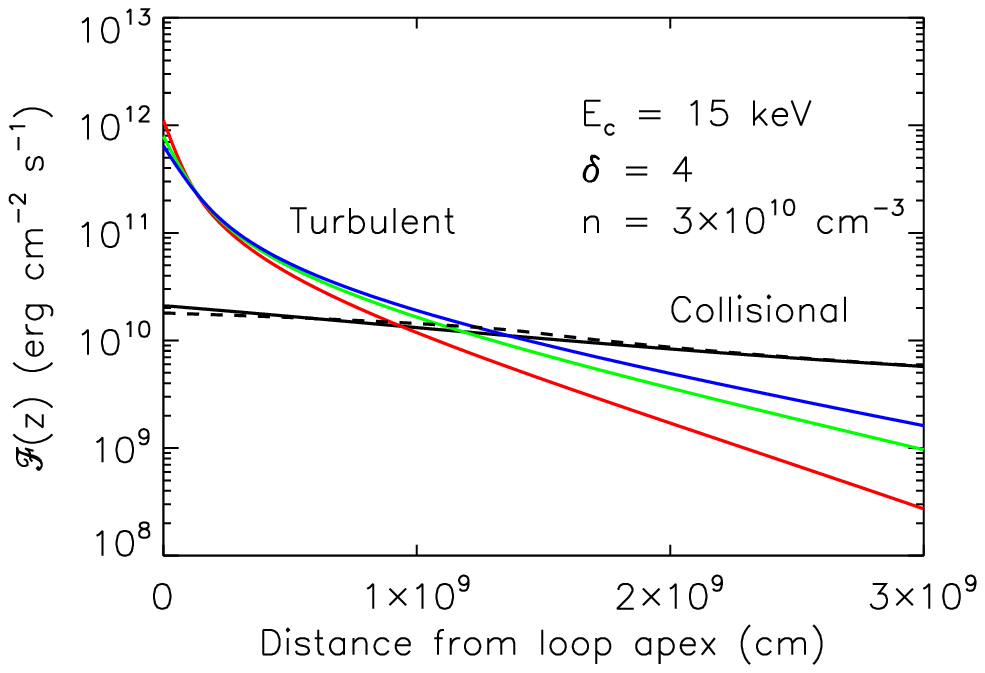}
\includegraphics[width=0.32\textwidth]{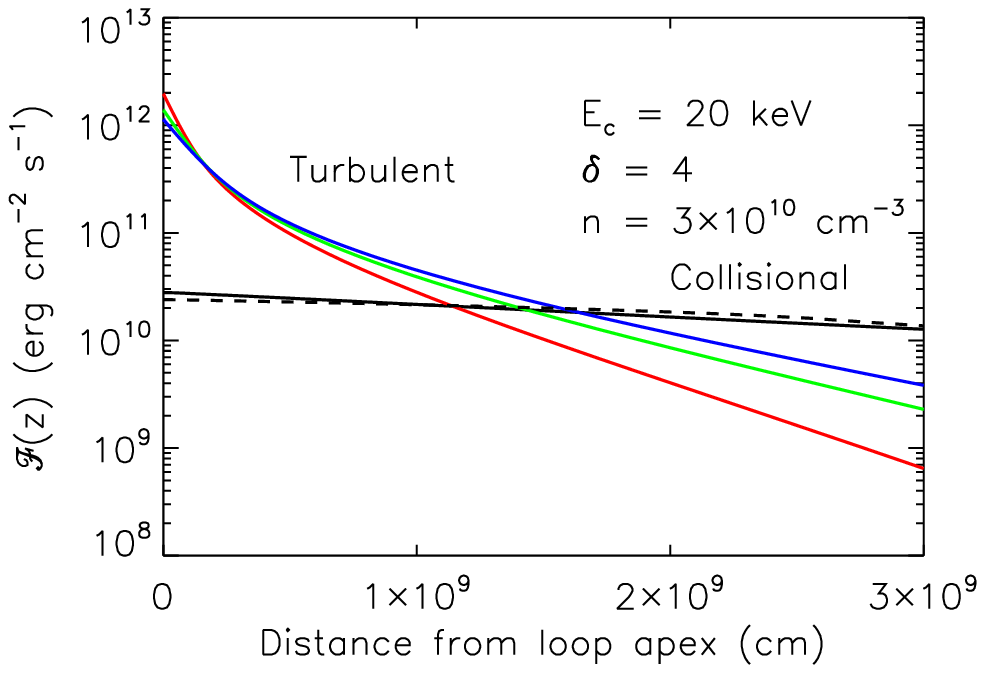}
\includegraphics[width=0.32\textwidth]{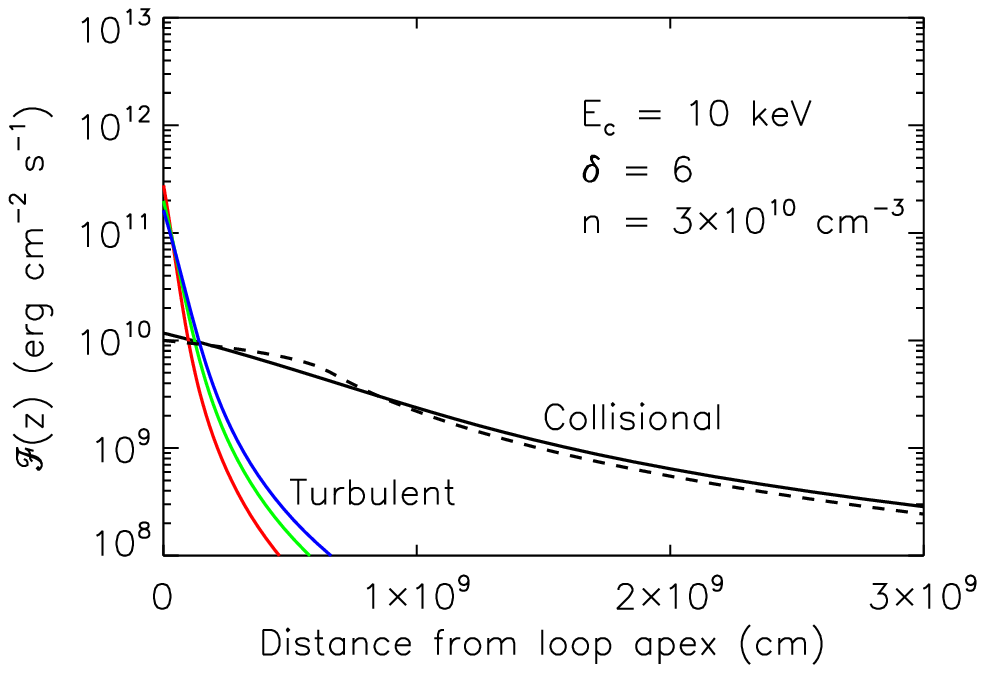}
\includegraphics[width=0.32\textwidth]{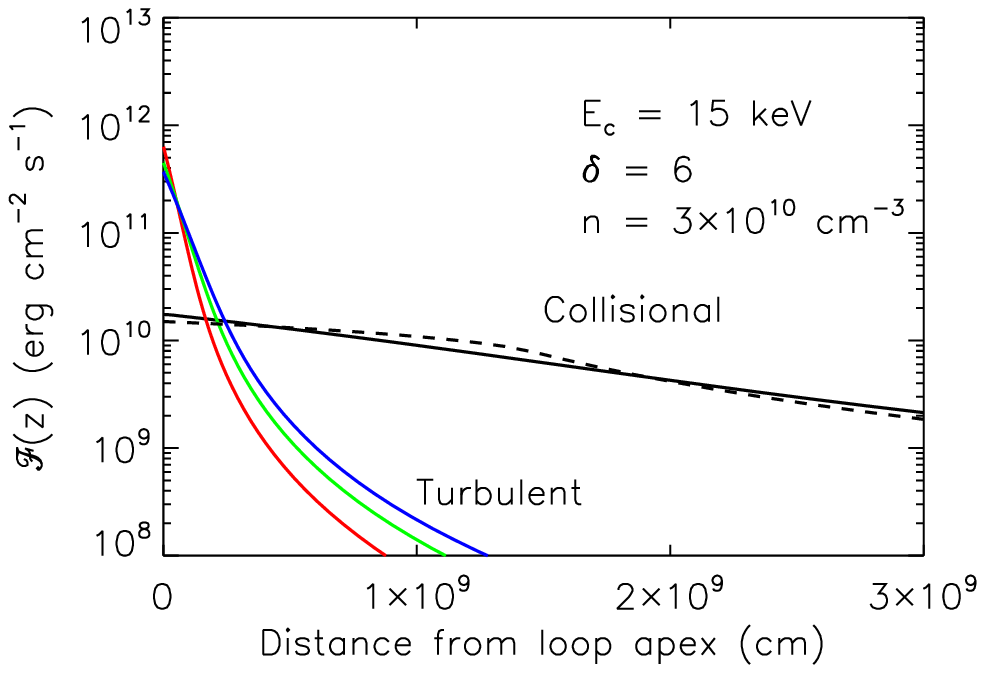}
\includegraphics[width=0.32\textwidth]{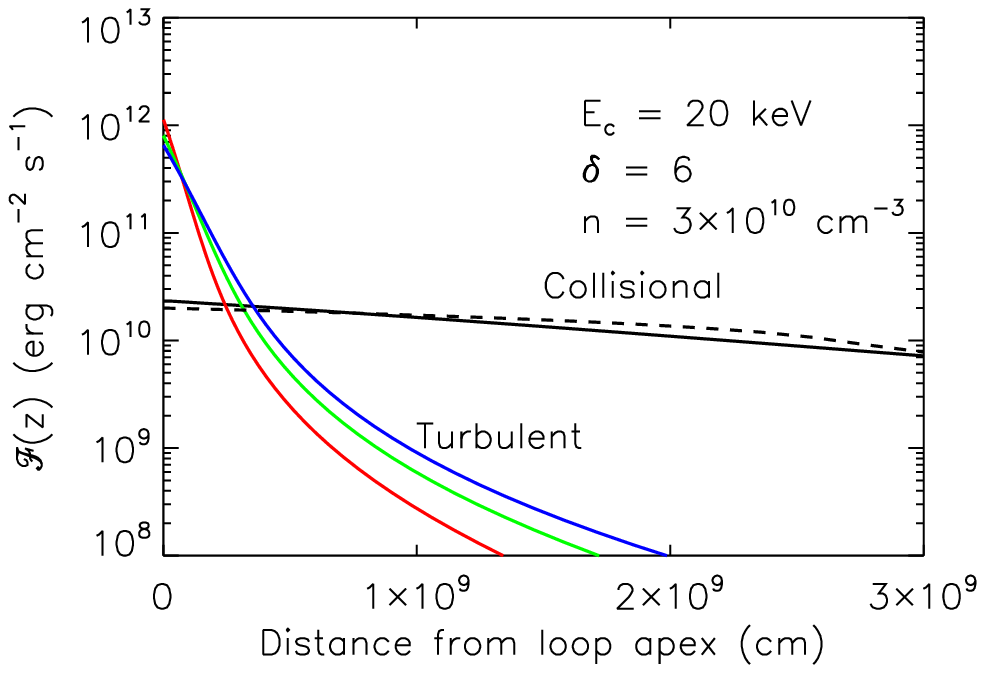}
\caption {Energy flux ${\cal F}(z)$ vs. distance $z$ from the (loop top) acceleration region. The solid line shows the results for collisional scattering, with the dashed line showing the behavior in the non-diffusive approach. The other lines show the results for turbulent scattering, with a turbulent scattering length $\lambda_T$ equal to $0.1 \times$ (red), $0.2 \times$ (green), and $0.3 \times$ (blue) the collisional mean free path of $7 \times 10^7$~cm.}\label{fig:energy_fluxes}
\end{center}
\end{figure}

Figure~\ref{fig:energy_fluxes} shows the spectrum-integrated energy flux (erg~cm$^{-2}$~s$^{-1}$) as a function of field-aligned distance $z$ from the beam injection point. Results are shown both for collisional scattering (solid line) and for turbulence-dominated scattering, using the three different values of the turbulent scattering mean free path discussed above: $\lambda_T = 7 \times 10^6, 1.5 \times 10^7$, and $2 \times 10^7$~cm. For comparison, the heating profile computed using the usual non-diffusive approach \citep[e.g.,][]{1972SoPh...26..441B,1973SoPh...31..143B,1978ApJ...224..241E} is shown as a dotted line. On the logarithmic scale used for the ordinate, the solid and dotted lines are almost indistinguishable \citep[cf. Figure~1 of][which is depicted using a linear scale for the ordinate]{2018ApJ...862..158E}. However, it is immediately evident that the behavior of energy flux vs. distance is markedly different for the turbulent scattering cases, with one-to-two orders of magnitude higher flux near the acceleration region and an even greater reduction in the energy flux at large distances, particularly for steeper spectra ($\delta=6$). The physical origin of this difference is the enhanced scattering rate of the accelerated electrons, in particular higher energy electrons, for which the turbulent scattering length $\lambda_T$ does not increase with energy (compared to the collisional scattering length which scales as $E^2$). This significantly different energy flux profile results in similarly large differences in the energy deposition profiles, as we next discuss.

\begin{figure}[pht]
\begin{center}
\includegraphics[width=0.32\textwidth]{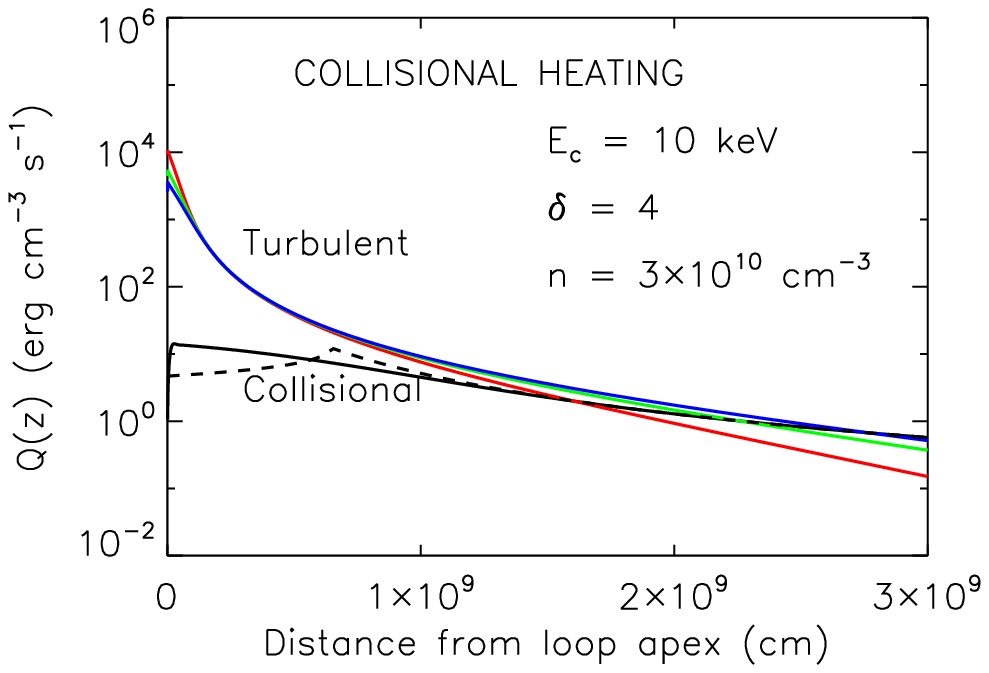}
\includegraphics[width=0.32\textwidth]{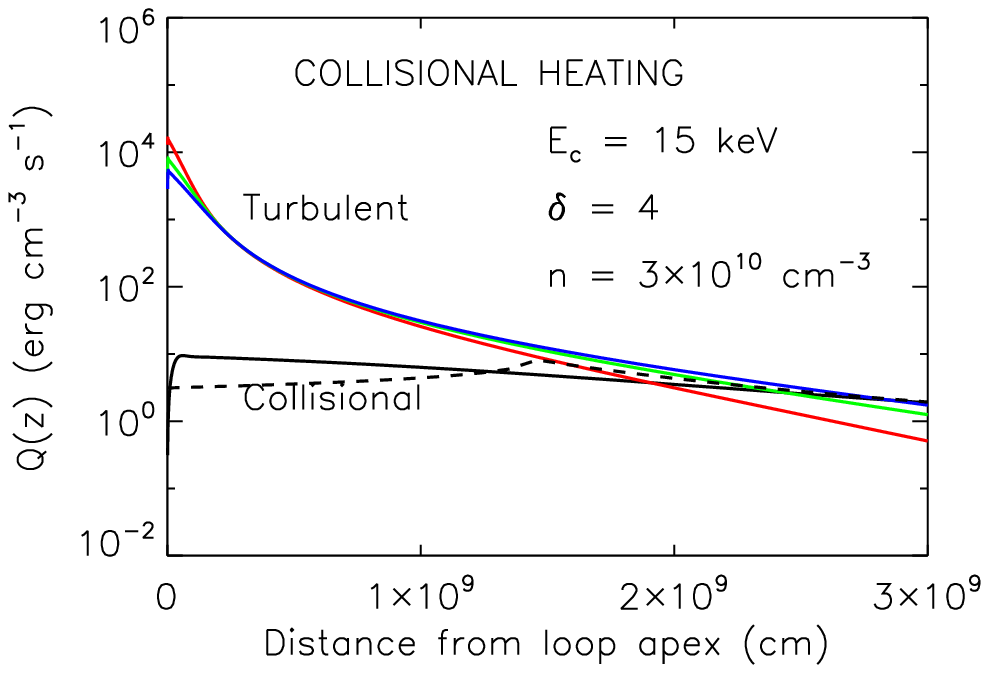}
\includegraphics[width=0.32\textwidth]{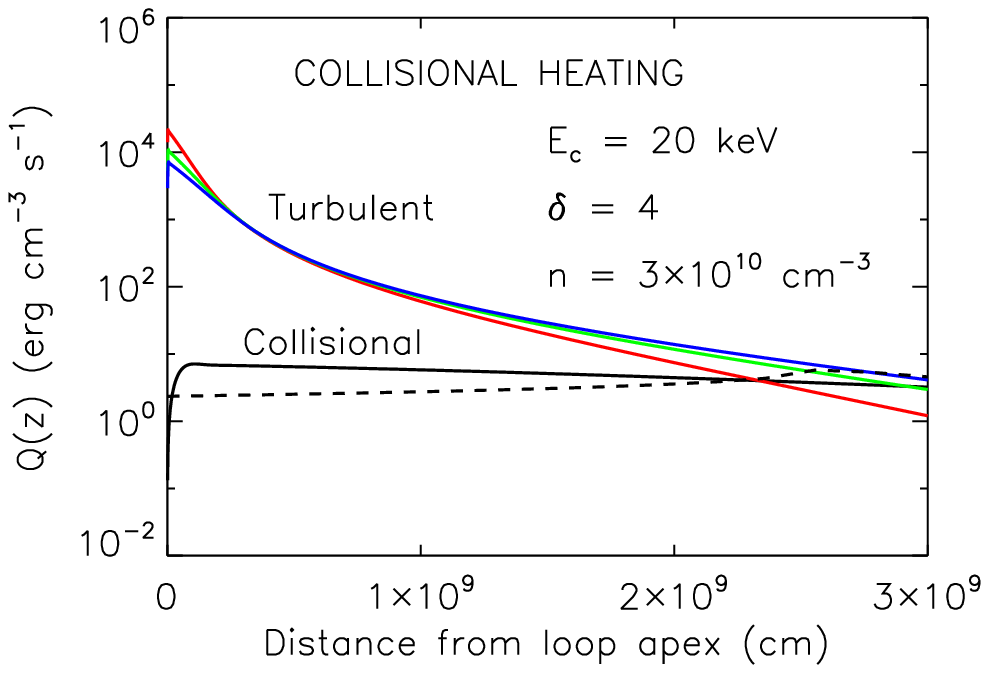}
\includegraphics[width=0.32\textwidth]{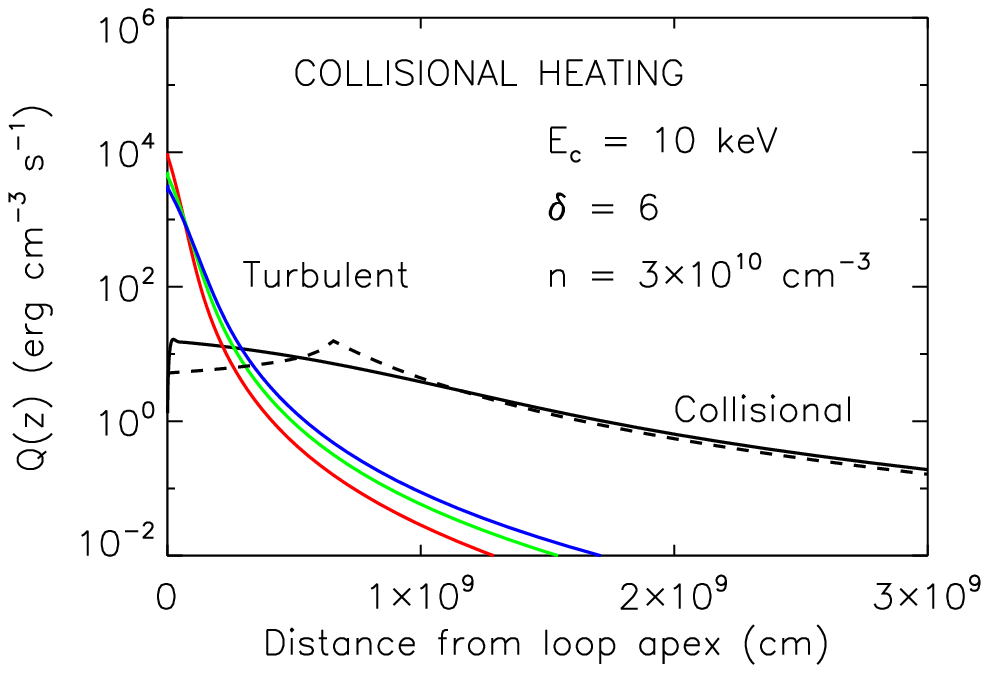}
\includegraphics[width=0.32\textwidth]{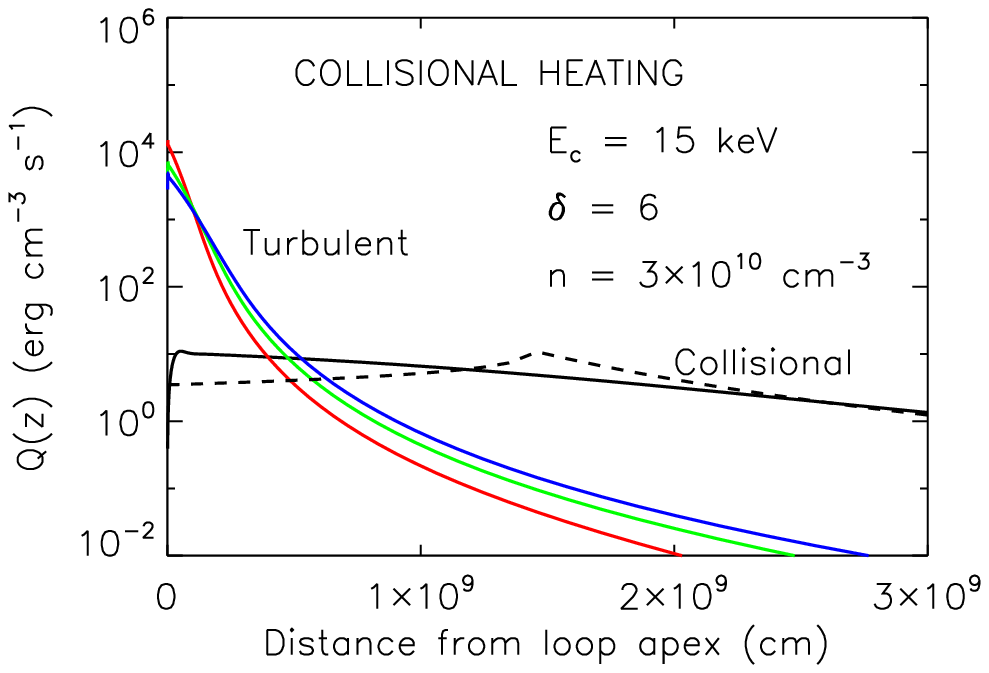}
\includegraphics[width=0.32\textwidth]{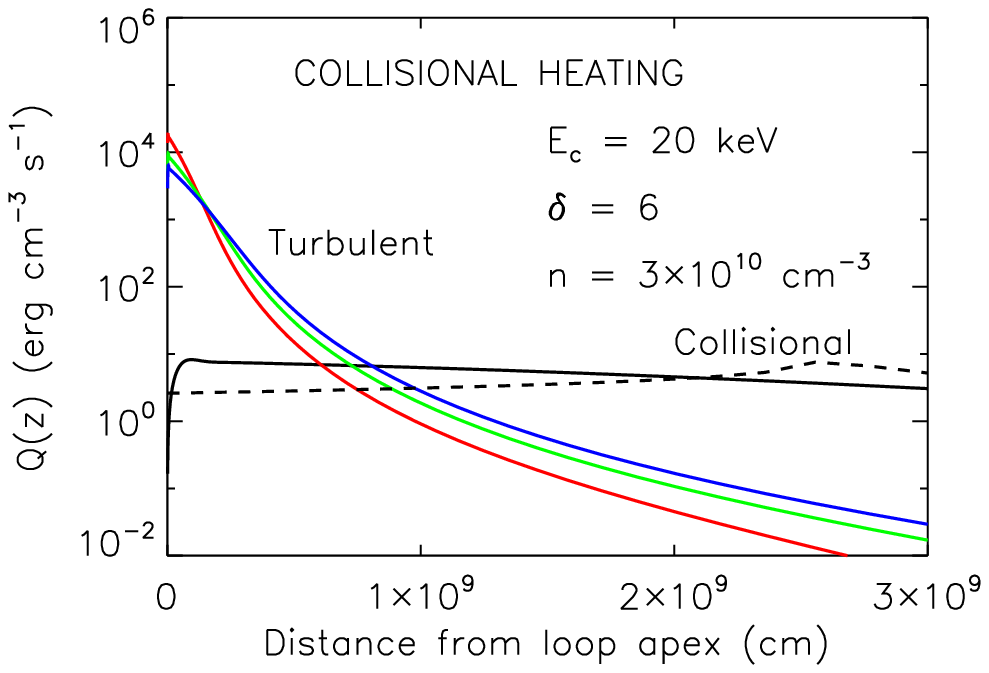}
\caption {Collisional heating rate $Q$ vs. distance from the (loop top) acceleration region. The solid line shows the results for collisional scattering,  with the dashed line showing the behavior in the non-diffusive approach. The other lines show the results for turbulent scattering, with a turbulent scattering length $\lambda_T$ equal to $0.1 \times$ (red), $0.2 \times$ (green), and $0.3 \times$ (blue) the collisional mean free path of $7 \times 10^7$~cm.}\label{fig:collisional-heating}
\end{center}
\end{figure}

Figure~\ref{fig:collisional-heating} shows the heating rates (erg~cm$^{-3}$~s$^{-1}$) due to collisions of beam electrons on ambient electrons. As in Figure~\ref{fig:energy_fluxes}, the dotted line shows the energy deposition rate in the commonly-used non-diffusive approach (note the local maximum for this calculation, a result of the sharp spectral break at $E_o=E_c$; the local maximum occurs at a distance $z \simeq E_c^2/2Kn \simeq 7 \times 10^6 \, E_C^2$~cm). The results for collisional diffusion differ from those of the non-diffusive model by a relatively modest factor of up to about two \citep[cf. Figure~1 of][noting the linear scale on the ordinate in that Figure]{2018ApJ...862..158E}, showing that the addition of scattering in a purely collisional analysis has a relatively small impact on the heating profile. However, a transport model that invokes turbulent diffusion results in much more significant changes to the heating profile.  In general the heating shifts significantly towards the top of the loop, with the energy deposition rate near the acceleration site enhanced by over an order of magnitude, and, particularly for steep injected spectra (see the $\delta=6$ cases), the energy deposition rate at large distances from the acceleration site reduced by several orders of magnitude. This much lower level of chromospheric heating leads to much lower pressures established in the chromosphere and, coupled with the very high pressures established by the strong heating in the corona, there should be significantly less pressure-gradient-driven ``evaporation'' of chromospheric material into the corona. This will be discussed further in Section~\ref{sec:discussion}.

\begin{figure}[pht]
\begin{center}
\includegraphics[width=0.32\textwidth]{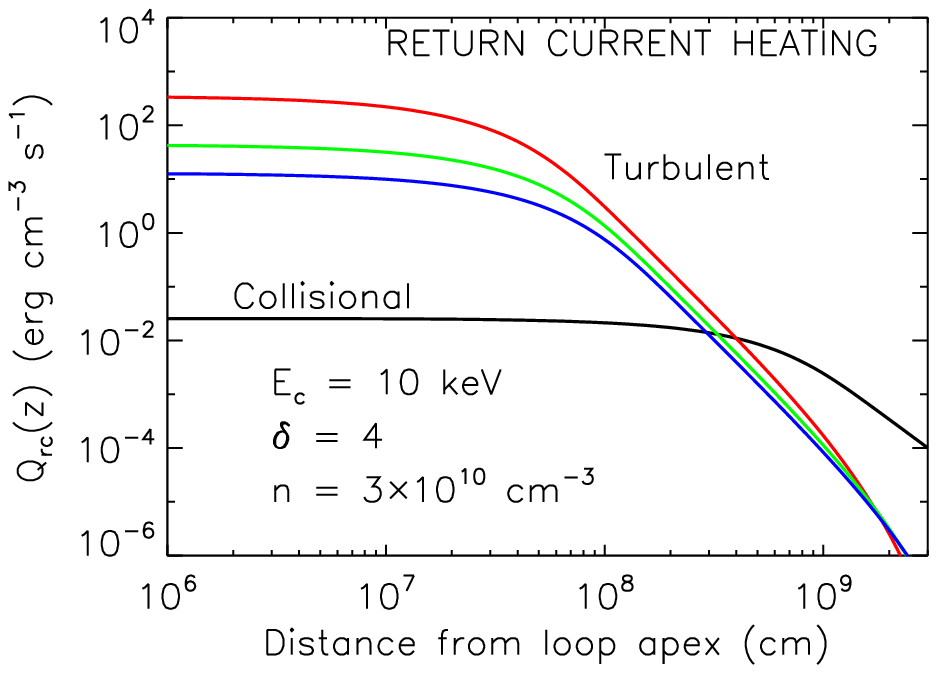}
\includegraphics[width=0.32\textwidth]{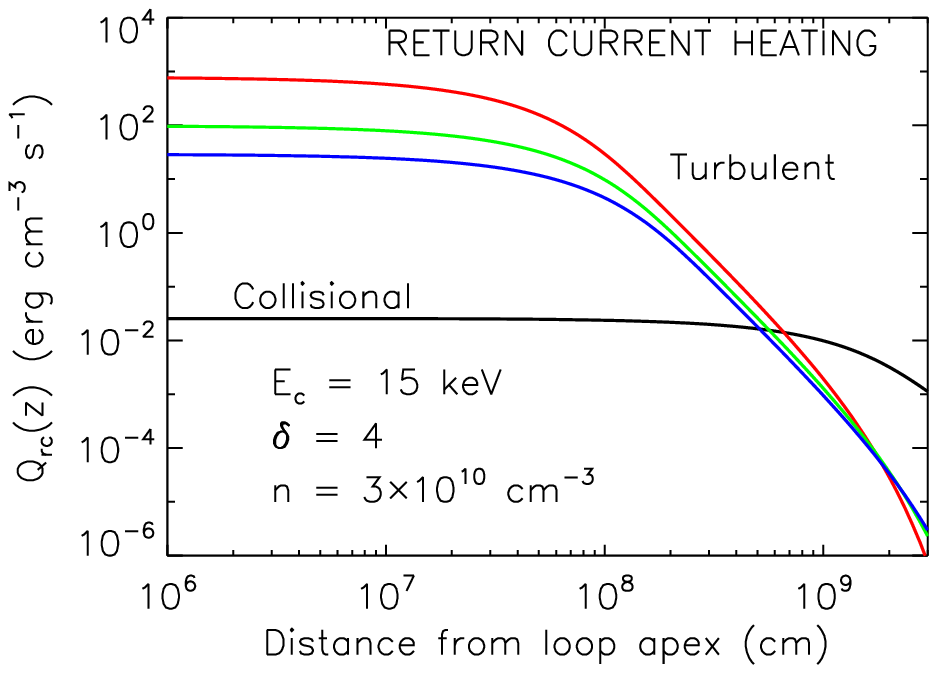}
\includegraphics[width=0.32\textwidth]{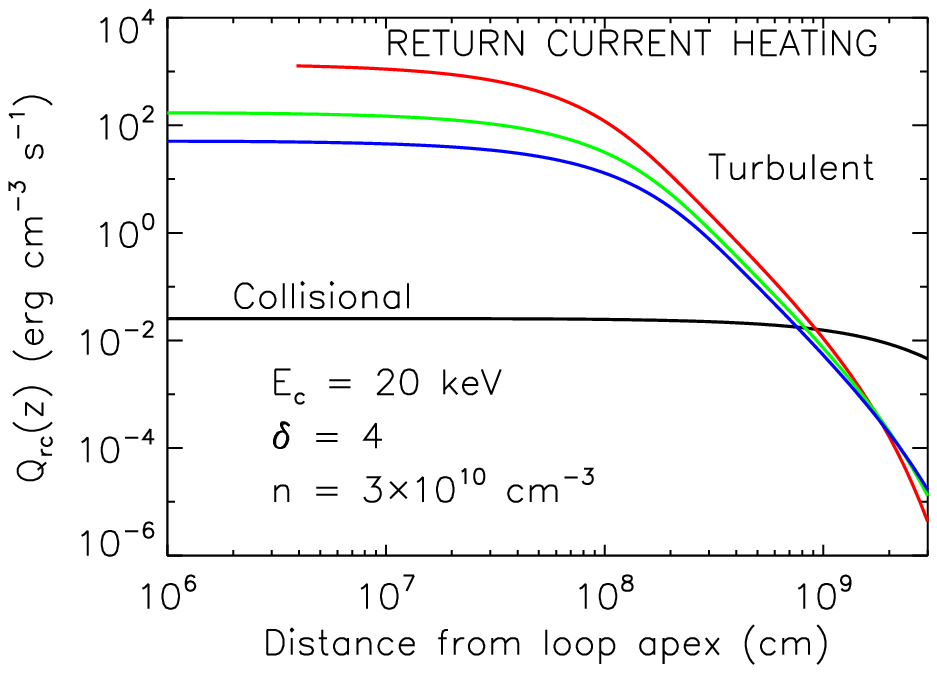}
\includegraphics[width=0.32\textwidth]{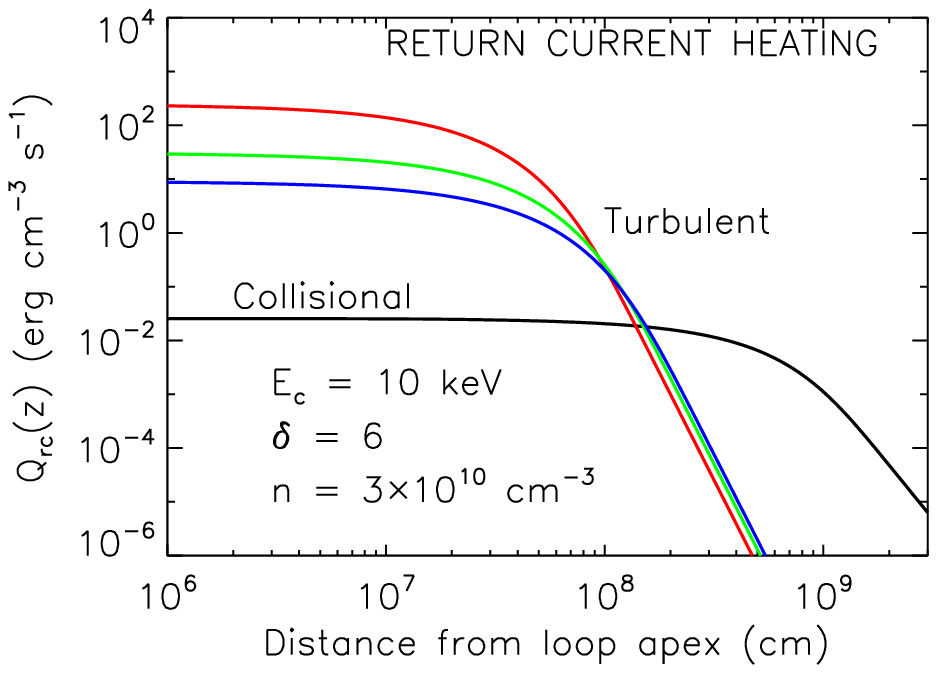}
\includegraphics[width=0.32\textwidth]{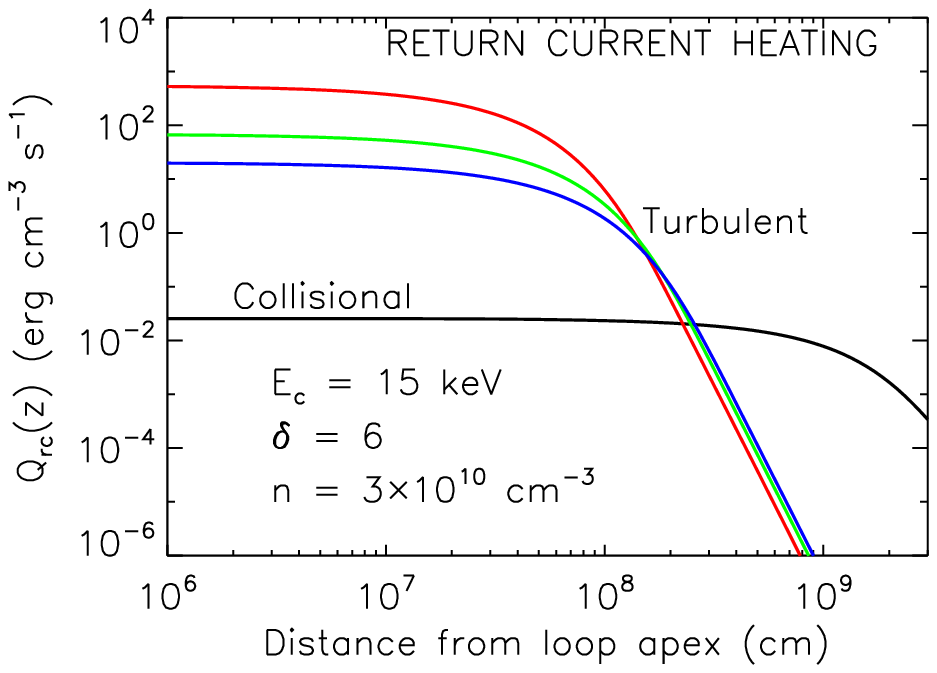}
\includegraphics[width=0.32\textwidth]{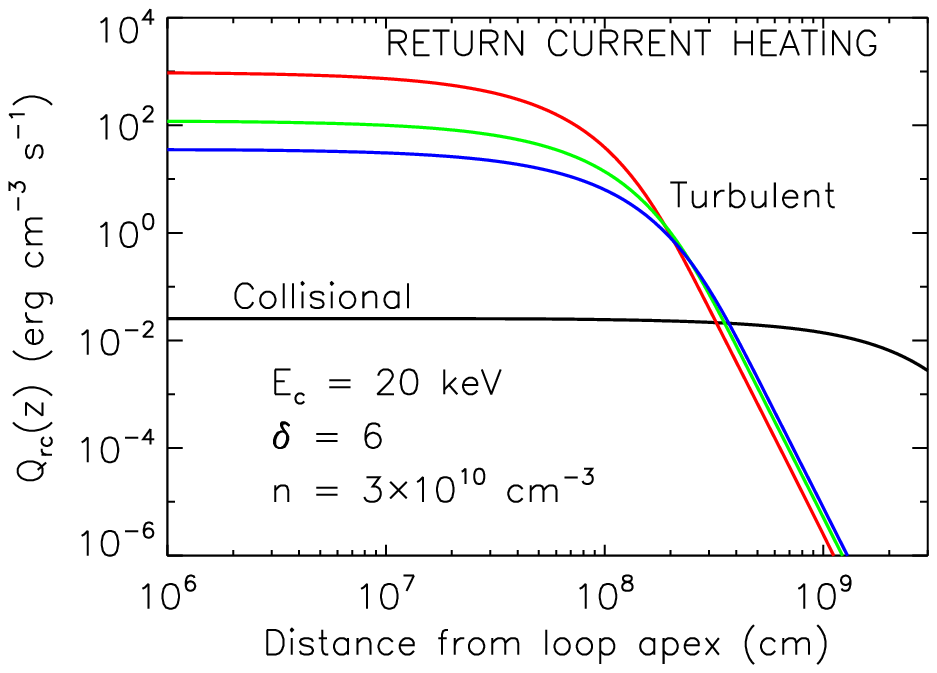}
\caption {Ohmic heating rate $Q_{rc}$ vs. distance from the (loop top) acceleration region. The solid line shows the results for collisional scattering, and the other lines show the results for turbulent scattering, with a turbulent scattering length $\lambda_T$ equal to $0.1 \times$ (red), $0.2 \times$ (green), and $0.3 \times$ (blue) the collisional mean free path of $7 \times 10^7$~cm. (Note that the abscissa is now on a logarithmic scale.)}\label{fig:rc-heating}
\end{center}
\end{figure}

\begin{figure}[pht]
\begin{center}
\includegraphics[width=0.32\textwidth]{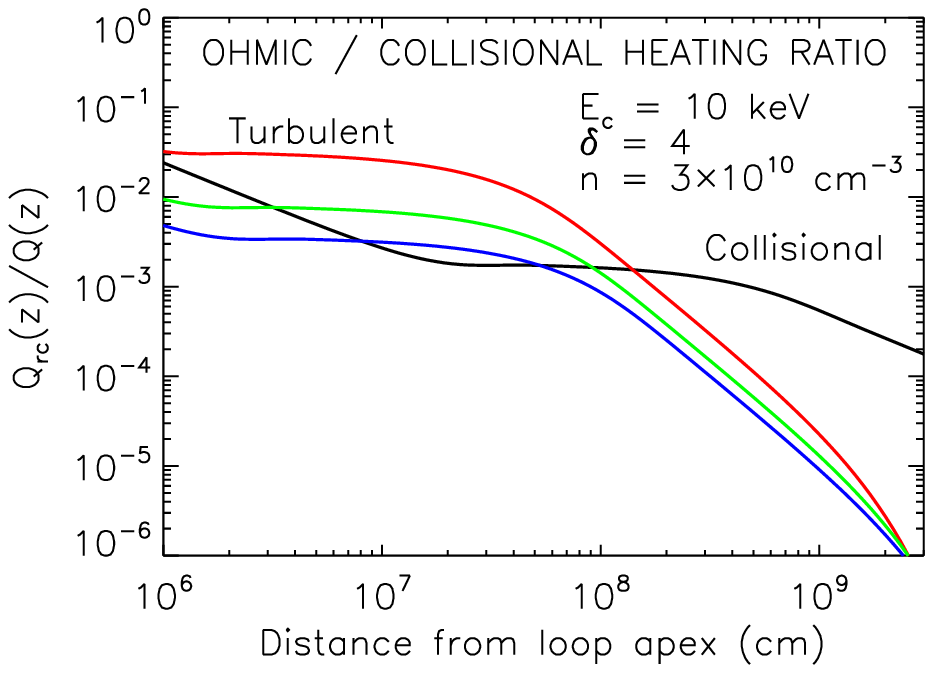}
\includegraphics[width=0.32\textwidth]{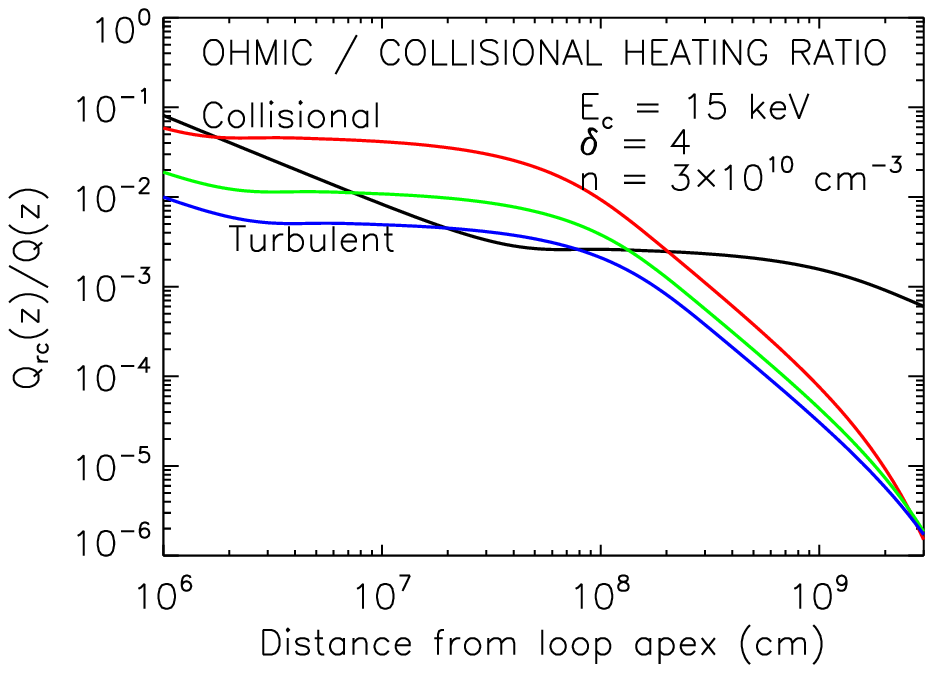}
\includegraphics[width=0.32\textwidth]{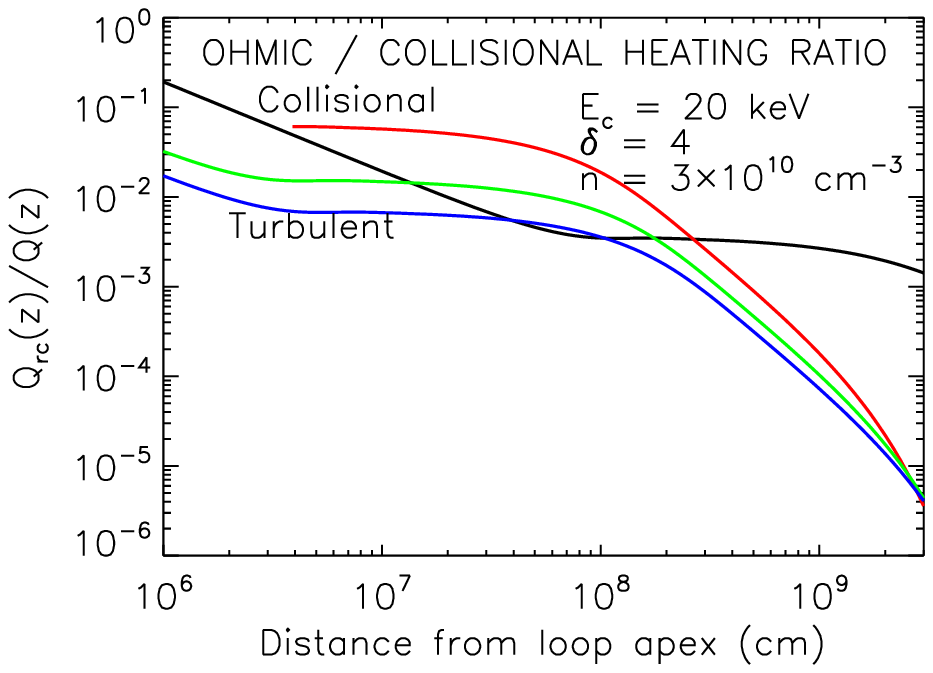}
\includegraphics[width=0.32\textwidth]{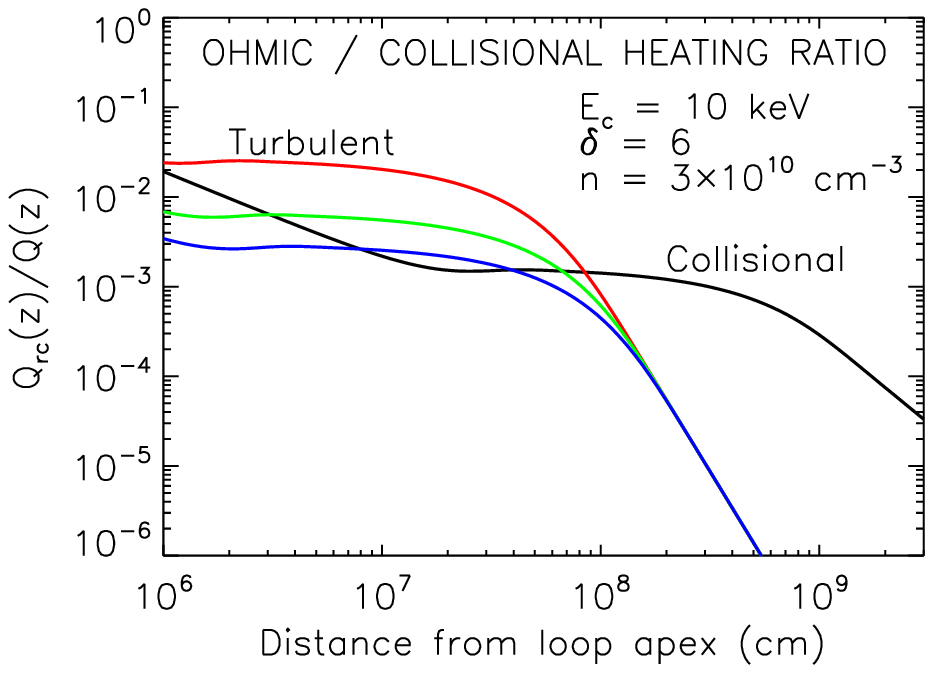}
\includegraphics[width=0.32\textwidth]{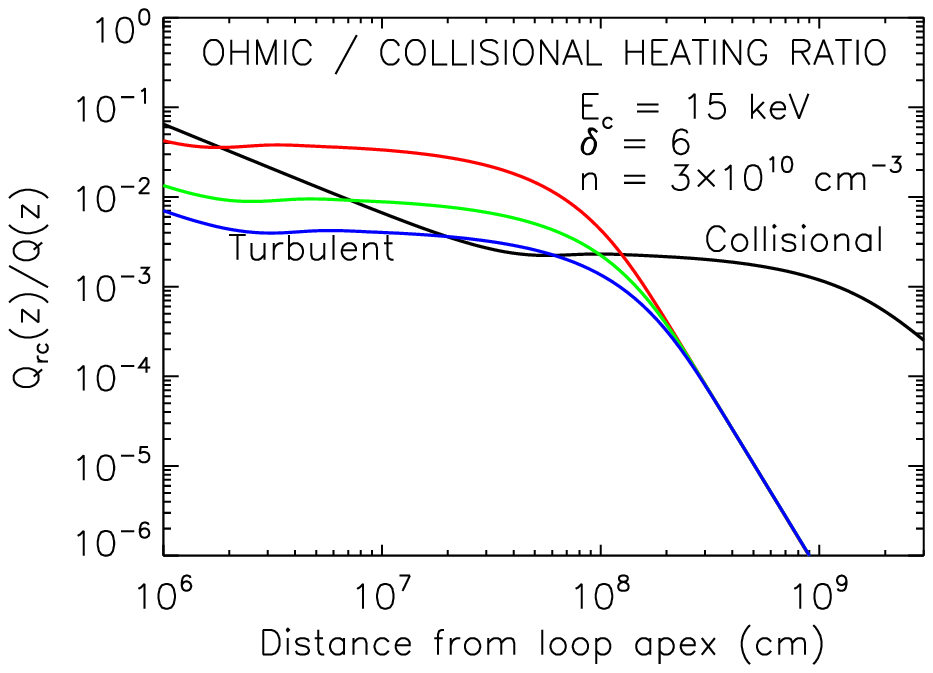}
\includegraphics[width=0.32\textwidth]{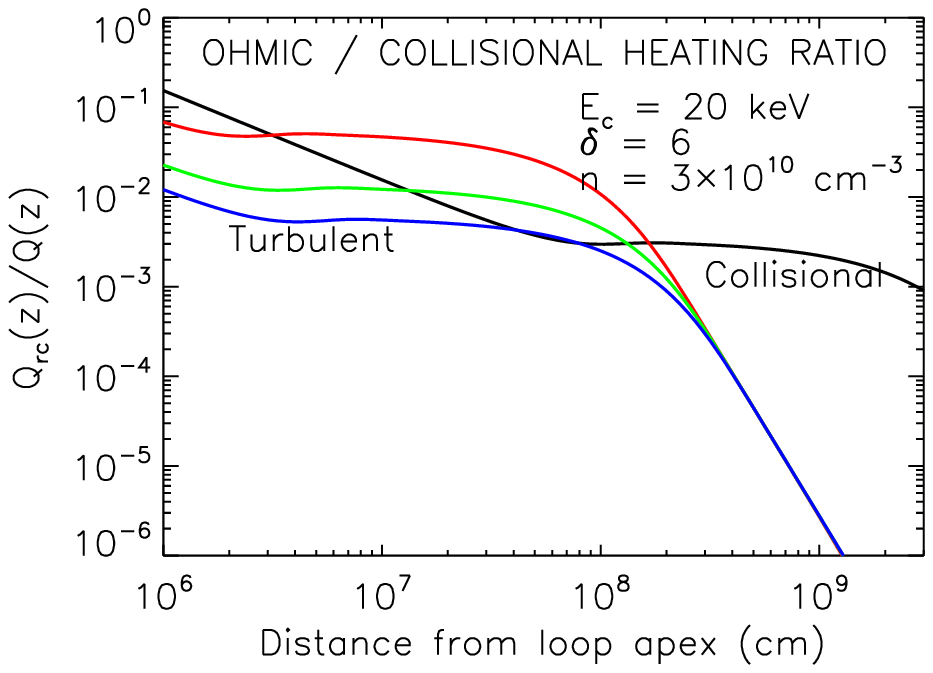}
\caption {Ratio of Ohmic to collisional heating $Q_{rc}/Q$. The solid line shows the results for collisional scattering, and the other lines show the results for turbulent scattering, with a turbulent scattering length $\lambda_T$ equal to $0.1 \times$ (red), $0.2 \times$ (green), and $0.3 \times$ (blue) the collisional mean free path of $7 \times 10^7$~cm. The ratio is $\ll 1$ at all positions in the target in all cases.}\label{fig:heating-ratio}
\end{center}
\end{figure}

Figure~\ref{fig:rc-heating} shows the heating rates due to Ohmic dissipation of the beam-neutralizing return current. Overall, the effect of enhanced scattering in the presence of strong turbulence is to reduce the anisotropy in the electron phase-space distribution function. This acts to reduce the magnitude of the net beam current and hence the magnitude of the return current density $j$. Because the amount of Ohmic heating scales as $j^2$, the Ohmic heating rate is reduced to a very small level, as evidenced by Figure~\ref{fig:heating-ratio}, which shows the ratio of the Ohmic heating rate to the heating effected by direct collisional heating. The ratio is everywhere $\ll 1$, for both collisional and turbulent scattering scenarios, even taking into account the enhanced Ohmic heating due to the increase in electrical resistivity in the turbulent case.  This justifies  {\it a posteriori} the use of a collisional treatment to derive the electron fluxes $F(E,z)$ used in calculating the Ohmic heating rate.

\section{Discussion and Conclusions}\label{sec:discussion}

The results show how the enhanced scattering of non-thermal electrons in the presence of strong turbulence has a very significant effect on the profile of electron energy flux vs. distance from the acceleration site, and hence on the spatial distribution of energy deposition by non-thermal electrons. Comparing results with the usually adopted test-particle treatment of electron beam transport and heating \citep[see, e.g.,][]{1984ApJ...279..896N,1989ApJ...341.1067M,2005ApJ...630..573A,2009ApJ...702.1553L,2022ApJ...931...60A} shows that the effects of a reduction in the scattering mean free path through turbulence are much more significant than the inclusion of scattering effects in a purely collisional analysis \citep{2018ApJ...862..158E}. The rapid isotropization of the electron distribution by turbulent scattering is consistent with analysis of photospheric albedo X-rays \citep[i.e, those that backscatter off the Sun;][]{2006ApJ...653L.149K,2013SoPh..284..405D,2024ApJ...964..145J} that suggest the electron distribution is often close to isotropic or only weakly anisotropic, a result that is inconsistent with field-aligned injection and a test-particle model for the evolution of the electron distribution.

The result has immediate consequences for studies in a number of areas, ranging from the importance of upward chromospheric evaporation flows driven by the pressure gradients established by the beam heating profile, downward chromospheric condensation flows driven by the rapid falloff in heating rate with depth in the upper chromosphere  \citep{1989ApJ...346.1019F} and for the resulting relationships between nonthermal and thermal emissions in the flare \citep{1985ApJ...289..414F,1987ApJ...312..895P,1990SoPh..129..113L,1991ApJ...381..572M,1993ApJ...413..786A,1995ApJ...444..478M,1996ApJ...459..804N,1999ApJ...521..906A,2004ApJ...611L..49W}. Specifically, the reduced level of chromospheric heating in models with strong turbulent scattering offers a compelling explanation for the fact that observed soft X-ray line profiles often do not show redshifts that are as significant as those expected from models that invoke collisional energy deposition by a stream of accelerated electrons that interact non-diffusively with the target atmosphere\citep[e.g.,][]{1986AdSpR...6f.159B,1989SoPh..123..161M}. Our results also have impact on studies of optical line intensities in flares \citep{2009A&A...499..923K}, coronal rain \citep{2020ApJ...890..100R}, stellar flares \citep{2003AdSpR..32.1057R,2007ASPC..362..304K} and even sunquakes \citep{2008SoPh..251..641Z}.

The enhanced coronal heating in models with strong turbulent scattering should result in higher coronal temperatures, which would act to offset the reduction in the thermal conductive flux effected by turbulent scattering \citep{2016ApJ...824...78B,2018ApJ...852..127B}.  \citep[On the other hand, the very significant reduction in return Ohmic current heating could help ameliorate the anomalously high coronal temperatures obtained in the return current modeling of][]{2024ApJ...977..246E,2025ApJ...993..127E}.
Enhanced coronal heating could also well account for the formation of looptop hard X-ray sources \citep{1995PASJ...47..677M,1998ApJ...505..418F,2008A&ARv..16..155K} and/or super-hot $>30$~MK thermal components in the decay phase of the flare \citep[e.g.][]{1985SoPh...99..263L}. However, the common presence of hard X-ray footpoints \citep[e.g.,][]{1995ApJ...454..522K,1995PASJ...47..355T,2002SoPh..210..307F,2007A&A...471..705J,2011ApJ...739...96K,2011ApJ...735...42B} does argue in favor of a significant electron flux reaching the chromosphere; either this places a lower limit on the value of the turbulence scattering length or suggests a distributed acceleration site lying along most of the flare loop, or electron re-acceleration \citep{2009A&A...508..993B} accomplished by acceleration of electrons due to wave-particle interactions in an inhomogeneous plasma \citep{2012A&A...539A..43K}.

In order to address this array of possible consequences, we strongly encourage the use of turbulence-scattering heating functions given by Equation~\eqref{eq:q-diff} in numerical codes that simulate the response of the solar atmosphere to impulsive phase energy input in a turbulent environment, especially when such turbulence is already incorporated in other terms in the energy equation, through turbulent modifications to the thermal and/or electrical conductivities.

\begin{acknowledgments}

We thank the referee for a careful reading of the paper and for valuable suggestions as to how improve it. A.G.E. was supported by NASA's Heliophysics Supporting Research Program through awards 80NSSC24K0244 and 80NSSC23K0448, by the NASA EPSCoR program through award number 80NSSC21M0362 to NASA Kentucky, and by the Kentucky Cabinet for Economic Development. E.P.K. acknowledges financial support from the STFC/UKRI grant ST/Y001834/1 and the Leverhulme Trust (Research Fellowship RF-2025-357). 

\end{acknowledgments}

\begin{contribution}

This work was initiated and developed by AGE, in consultation with EPK. Both authors contributed to the writing of the manuscript and agree with its contents.

\end{contribution}

\bibliography{refs}{}
\bibliographystyle{aasjournalv7}

\end{document}